\documentclass[letterpaper, 10 pt, conference]{ieeeconf}  

\IEEEoverridecommandlockouts                              
\usepackage{amssymb}
\usepackage{graphics}
\usepackage{epsfig}
\usepackage{mathptmx}
\usepackage{times}
\usepackage{amsmath}

\newtheorem{example}{Example}

\title{\LARGE \bf Hybrid Systems and Control With Fractional Dynamics (II): Control}
\author{S. Hassan~HosseinNia, In\'{e}s~Tejado, and~Blas~M.~Vinagre
\thanks{This work was supported by the Spanish Ministry of Economy and Competitiveness under 
the research project DPI2012-37062-C02-02.}
\thanks{S. H. HosseinNia, I. Tejado and B. M. Vinagre are with Department of Electrical, Electronic and Automation Engineering, Industrial Engineering School, University of Extremadura, 06006 Badajoz, Spain. e-mail: \tt \{hoseinnia,itejbal,bvinagre\}@unex.es}}

\begin{document}

\maketitle
\thispagestyle{empty}
\pagestyle{empty}

\begin{abstract}
No mixed research of hybrid and fractional-order systems into a cohesive and multifaceted whole can be found in the literature. This paper focuses on such a synergistic approach of the theories of both branches, which is believed to give additional flexibility and help the system designer. It is part II of two companion papers and focuses on fractional-order hybrid control. Specifically, two types of such techniques are reviewed, including robust control of switching systems and different strategies of reset control. Simulations and experimental results are given to show the effectiveness of the proposed strategies. Part I will introduce the fundamentals of fractional-order hybrid systems, in particular, modelling and stability of two kinds of such systems, i.e., fractional-order switching and reset control systems. 
\end{abstract}

\section{Introduction}
Hybrid systems (HS) are heterogeneous dynamic systems whose behaviour is determined by interacting continuous-variable and discrete-event dynamics, and they arise from the use of finite-state logic to govern continuous physical processes or from topological and networks constraints interacting with continuous control \cite{Gollu_89,schumacher_99,Goebel_09}. It is worth mentioning that, among them, we focuses on two kinds of HS in this work: switching and reset control systems. Switching systems, a class of HS consisting of several subsystems and a switching rule indicating the active subsystem at each instant of time, have been the subject of interest for the past decades, for their wide application areas. Likewise, reset control systems are standard control systems endowed with a reset mechanism, i.e., a strategy that resets to zero the controller state (or part of it) when some condition holds. The hybrid behaviour comes from the instantaneous jump due to resets of whole or part of system states \cite{banos2011,Schutter2009}.

Many real dynamic systems are better characterized using a fractional-order dynamic model based on differentiation and integration of non-integer-order. The concept of fractional calculus has tremendous potential to change the way we see, model, and control the nature around us. Denying fractional derivatives is like saying that zero, fractional, or irrational numbers do not exist. From the control engineering point of view, improving and developing the control is the major concern (see e.g. \cite{Monje10,Podlubny_99a}). 

Recently, the wide applicability of both HS and systems with fractional-order dynamics has inspired a great deal of research and interest in both fields. Unfortunately, in general there are many difficulties in mixing different mathematical domains. The case of combining the theories of such systems is no exception. Given this motivation, this paper arises from the idea of coupling two different distinct branches of research, fractional calculus and HS, into a synergistic way, which is believed to give additional flexibility and help the system designer, taking advantage of the potentialities of both worlds. To this respect, a mathematical framework of fractional-order hybrid systems (FHS), including modeling, stability analysis, control and simulation, is required to be developed. Accordingly, part II of these two companion papers deals with fractional-order hybrid control. In particular, two types of such techniques are reviewed, robust control of switching systems and different reset control strategies, and analysed using the theory developed for FHS in part I \cite{HosseinNia2014a}. Experimental and simulated examples are given to demonstrate the effectiveness of the proposed strategies. 

The remainder of part II of this paper is organized as follows. Sections~\ref{Robust} and \ref{ADRSC} deal with robust control for switching systems and fractional-order reset control, respectively, as particular cases of fractional-order hybrid control systems. Some simulation and experimental applications are given in each section. Concluding remarks are included in Section~\ref{Conch2}.

\section{Robust Fractional-Order Control of Switching Systems}
\label{Robust}

This section addresses the main issues involved in a frequency-domain design method for switching systems for both integer- or fractional-order controllers, taking into account specifications regarding performance and robustness and ensuring the stability of the controlled system. The velocity control of a vehicle given two design specifications is shown as an example of application. The full description of this control technique can be found in \cite{Hosseinnia2013}. 

\begin{figure}[ht]
\begin{center}
\includegraphics[width=0.4\textwidth]{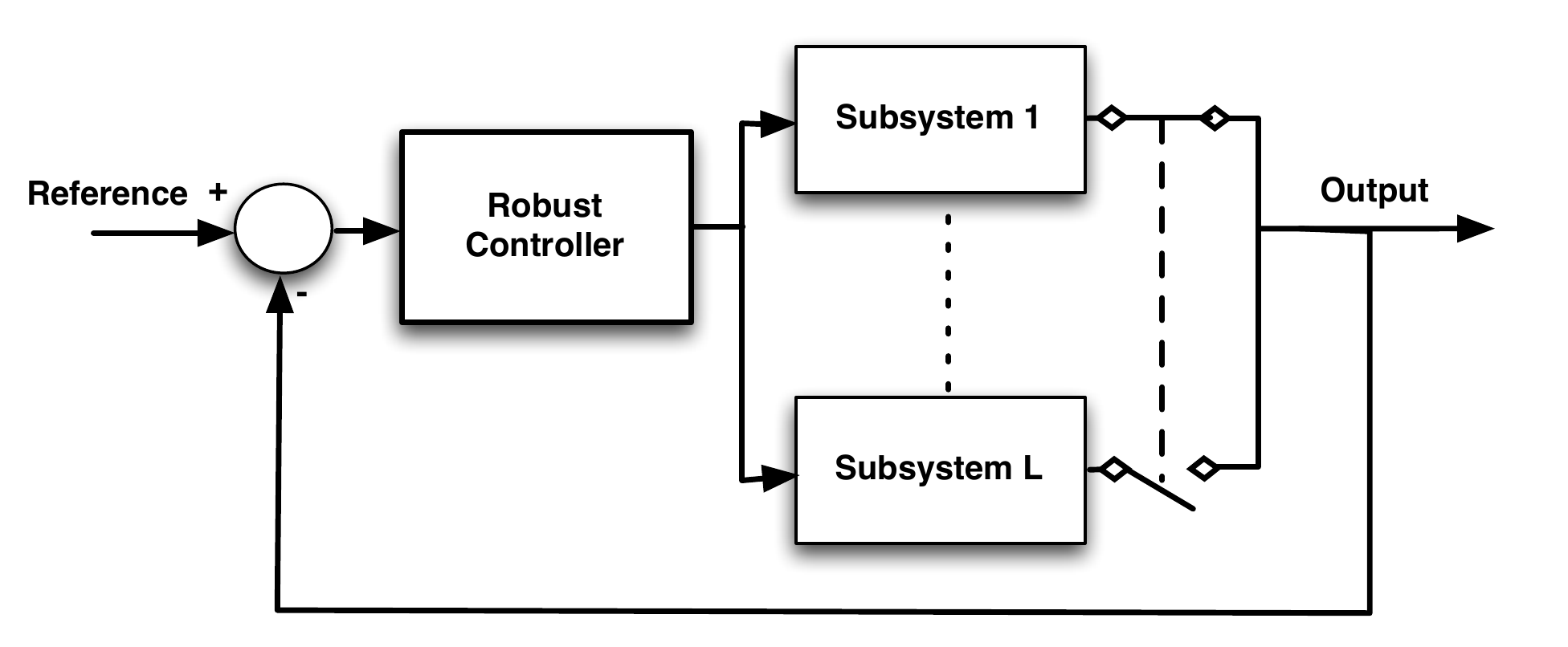}
\end{center}
\caption{Scheme of the controlled system}%
\label{RobustPI}%
\end{figure}

A scheme of the strategy is shown in Fig.~\ref{RobustPI} for a general system with subsystems $G_i(s)$, with $i=1,2,...,L$ subsystems. Specifications related to phase margin, gain crossover frequency and output disturbance rejection are going to be considered in this design method. Indeed, other kinds of specifications can be met, depending on the particular requirements of the application. It should be noticed that, apart from these design specifications, which can change with the application, the stability conditions have to be also fulfilled. Actually, if the number of subsystems which constitutes the system to be controlled is $L$, there are $L-1$ stability conditions to be fulfilled. Therefore, denoting the number of specifications as $N$, a controller with $L+N-1$ parameters is required in order to fulfil all given specifications and the stability conditions. 

Consider $G_n$ as the subsystem with the worst conditions for each specification and assume that the phase margin and gain crossover frequency of such a subsystem are denoted as $\phi_{m_n}$ and $\omega_{cp_n}$, respectively. Also consider that $c_{i}$, $i=1,2,...,L$, are the characteristic polynomials of each closed-loop subsystem and $K(j\omega)$ is the controller to be tuned. Thus, the design problem can be formulated as follows:
\begin{enumerate}
\item Frequency domain specifications:
\begin{enumerate}
\item Phase margin:
\begin{equation}
\arg(K(j\omega_{cp_{n}})G_{n}(j\omega_{cp_{n}}))+\pi>\phi_{m_n}.\label{SPe1}
\end{equation}
\item Gain crossover frequency:
\begin{equation}
\left|  K(j\omega_{cp_{n}})G_{n}(j\omega_{cp_{n}}) \right|  _{dB}=0 dB. \label{SPe2}%
\end{equation}
\item Output disturbance rejection:
\small
\begin{equation}
 \left\vert S(j\omega)=\frac{1}{1+G_n(j\omega)K(j\omega)} \right\vert_{dB} \leq M, \forall\omega \leq \omega_s, \label{SPe3}%
\end{equation}
\normalsize
where $M$ is the desired value of the sensitivity function $S$ for frequencies less than $\omega_s$.
\end{enumerate}
\item Stability conditions:
\begin{eqnarray}
\nonumber \left\vert \arg(c_{1}(j\omega))-\arg(c_{2}(j\omega))\right\vert <\frac{\pi}%
{2},\forall\omega \geq 0, \\
\nonumber \vdots\\
\left\vert \arg(c_{L-1}(j\omega))-\arg(c_{L}(j\omega))\right\vert <\frac{\pi}%
{2},\forall\omega \geq 0.
\label{rL}
\end{eqnarray}
\end{enumerate}

It is important to remark that (\ref{SPe1})-(\ref{SPe3}) refer to the worst conditions concerning phase margin, crossover frequency and sensitivity for a subsystem $G_n$ among all subsystems. The same can be done for any other specification, such as, high frequency noise rejection, steady-state error cancellation, etc. (see e.g.~\cite{Monje08} for more tuning specifications). The set of conditions (\ref{rL}) ensures the quadratic stability of the switching system. In the case of the fractional-order systems or time delayed systems, an approximation of fractional-order derivative or delay can be used to apply these specifications. 

As an example, in the case of $N=2$, a list of types of switching systems and their possible controllers is given in Table~\ref{MultiP} --any other type of controller can be tuned using the same idea. As can be stated, the use of fractional-order controllers may have the advantage of allowing more specifications or subsystems to be fulfilled or controlled, respectively, and, consequently, more robust performances to be attained.

\begin{table}[ht]
\caption{Type of switching system and possible controllers when $N=2$}
\begin{center}
\begin{tabular}{ l | c | c }
$L$ & Type of controller & Transfer function\\ \hline
$2$ & PID & $K_p+\frac{K_i}{s}+K_ds$ \\ \hline
$2$ & Fractional PI (FPI) & $K_p+\frac{K_i}{s^\lambda}$ \\ \hline
$2$ & Fractional PD (FPD) & $K_p+K_ds^\mu$ \\ \hline
$3$ & PID with noise filter (NPID)& $K_p+\frac{K_i}{s}+\frac{K_ds}{1+s/NN}$ \\ \hline
$4$ & Fractional PID (FPID) & $K_p+\frac{K_i}{s^\lambda}+K_ds^\mu$ 
\end{tabular}
\end{center}
 \label{MultiP}   
 \end{table}

To determine the controller parameters, the set of nonlinear equations (\ref{SPe1})--(\ref{rL}) has to be solved. To do so, the optimization toolbox of Matlab can be used to reach out the best solution with the minimum error. More precisely, the function \textit{FMINCON} is able to find the constrained minimum of a function of several variables. It solves problems of the form $\min_{x}f(x)$ subject to: $C(x) \leq 0$, $C_{eq}(x)=0$, $x_m
\leq x\leq x_M$, where $f(x)$ is the function to minimize; $C(x)$ and
$C_{eq} (x)$ represent the nonlinear inequalities and equalities, respectively
(non-linear constraints); $x$ is the minimum we are looking for; and $x_m$ and
$x_M$ define a set of lower and upper bounds on the design variables, $x$. 

In this particular case, the specification (\ref{SPe1}) will be taken as the main function to minimize, and the rest of specifications, i.e., (\ref{SPe2})-(\ref{rL}), will be taken as constrains for the minimization, all of them subjected to the optimization parameters defined within the function \textit{FMINCON}. The success of this optimization process depends mainly on the initial conditions considered for the parameters of the controller.

\begin{example} Velocity control of a vehicle with first-order dynamics given two design specifications.
\label{Excar}
\end{example}
In \cite{HosseinNia2011,HosseinNia_12}, we proposed a hybrid
model of a vehicle taking into account its different dynamics when accelerating and braking as follows
\begin{eqnarray}
\label{throttle}
G_{1}(s)\simeq\frac{4.39}{s+0.1746},\\ %
\label{brake_eq}
G_{2}(s)\simeq\frac{4.45}{s+0.445}, %
\end{eqnarray}
where $G_{1}$ and $G_{2}$ refer to the throttle and brake dynamics, respectively. The input and the output of the system are the reference and the actual velocities of the car.

From the viewpoint of the comfort of the car's occupants, phase margin and crossover frequency has to be chosen around $80^\circ$ and $0.8$ rad/s, respectively, in order to obtain a smooth closed-loop response with an overshoot close to $0$. Therefore, given two specifications, $N=2$, and two subsystems, $L=2$, controllers with three parameters are required to this application. In particular, two different three-parameter controllers are designed: a fractional PI (FPI) and a traditional PID controllers of the forms given in Table~\ref{MultiP}. Solving the set of equations (\ref{SPe1})--(\ref{rL}) for the previous specifications, the parameters of both controllers are: 
\begin{enumerate}
\item FPI: $K_{p_1}=0.15$, $K_{i_1}=0.07$, $\alpha=0.71$; 
\item PID: $K_{p_2}=0.1$, $K_{i_2}=0.11$ and $K_d=0.223$.
\end{enumerate}

\begin{figure}[ht]
\centering
\begin{tabular}{c}
\small
($a$)\\
\normalsize
\includegraphics[width=0.4\textwidth] {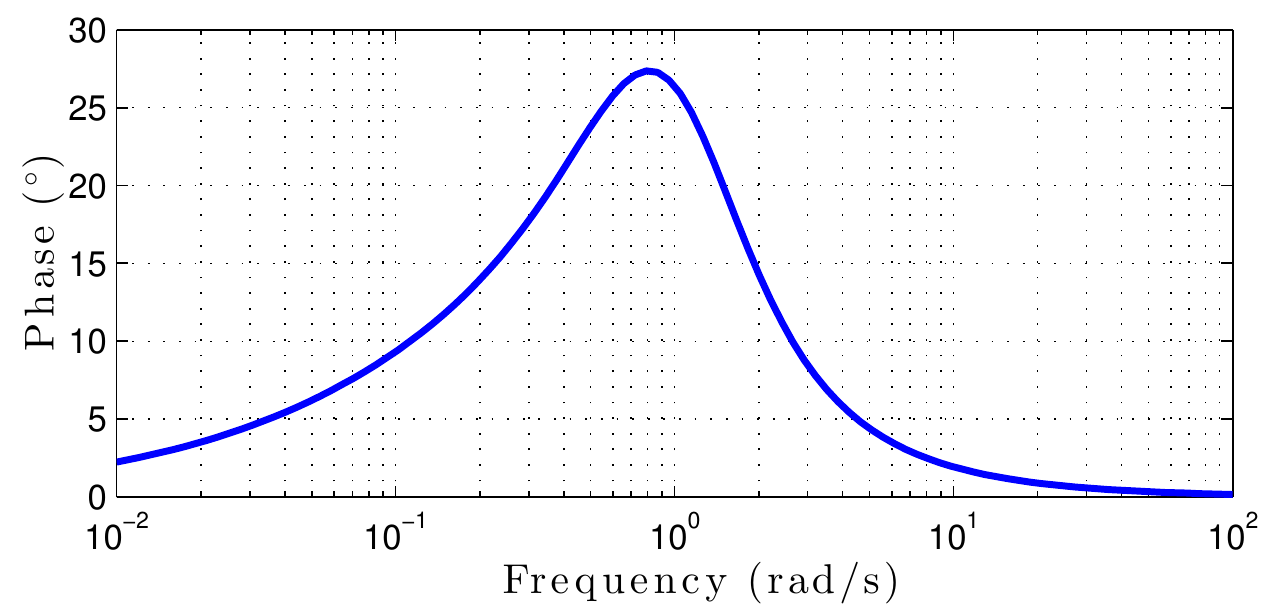} \\
\small
($b$)\\
\normalsize
\includegraphics[width=0.4\textwidth] {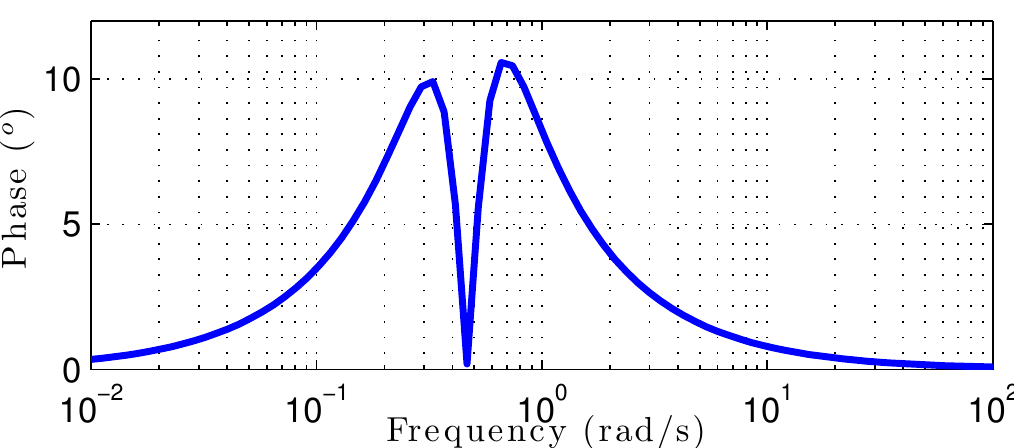}
\end{tabular}
\caption{Phase difference between the two characteristic polynomials of the closed-loop system in Example~\ref{Excar} when applying: ($a$) FPI ($b$) PID}
\label{phase}
\end{figure}

The phase difference between the two characteristic polynomials of the closed-loop controlled subsystems for both cases is shown in Fig.~\ref{phase}. It is observed that the maximum phase differences are $27.35$ and $10.57^\circ$ when using the FPI and PID, respectively, so the controlled system is quadratically stable in both cases. 

\begin{figure}[ptbh]
\begin{center}
\includegraphics[width=0.375\textwidth]{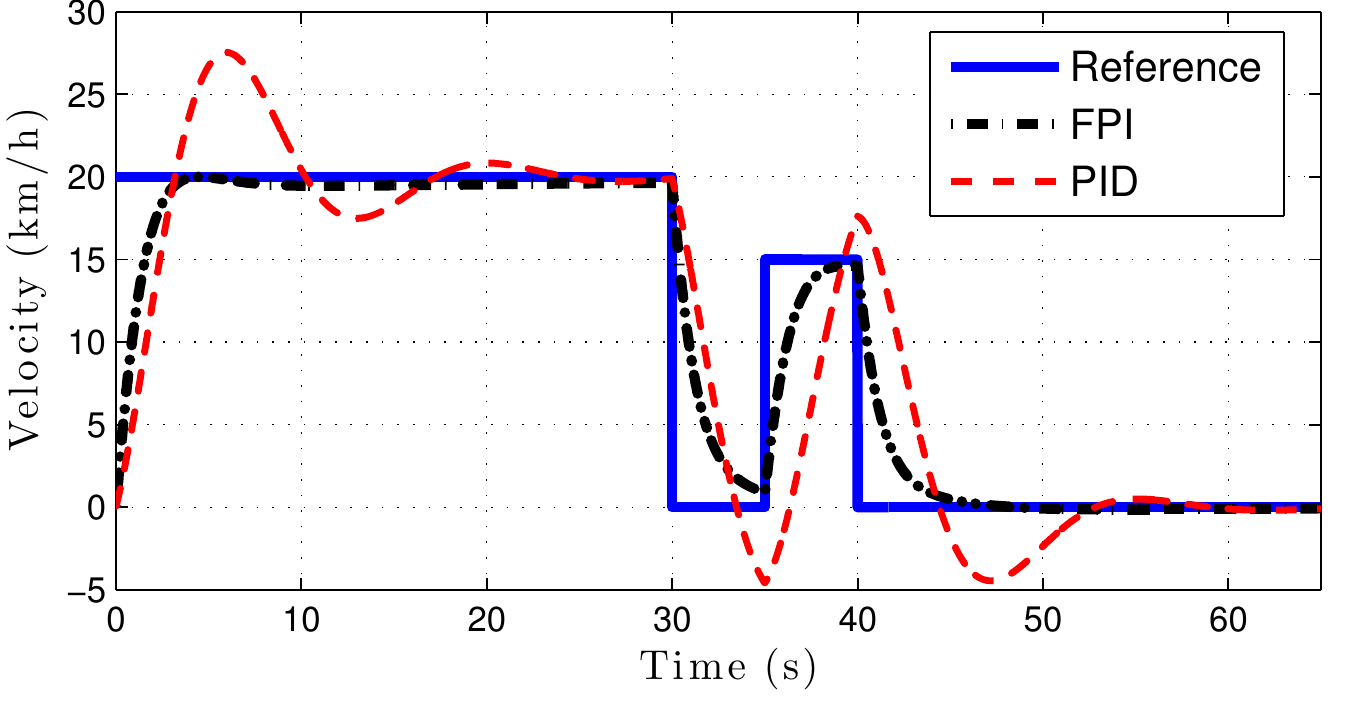}
\end{center}
\caption{Time response of the controlled system in Example~\ref{Excar} with FPI and PID during random switching}%
\label{Exp1comp}%
\end{figure}

To show the system performance in time domain, a manoeuvre which simulates the increase and the decrease of the car velocity from or to $0$ km/h --stop completely-- during random switching is depicted in Fig.~\ref{Exp1comp} for the FPI and PID cases. It can be observed that  the car has an adequate performance for both the throttle and the brake actions when applying the FPI controller (dash-dotted black line), achieving the reference velocity in a suitable time and without overshoot in both cases. Although both controllers fulfilled the specifications, the response when using the PID (dashed red line) has a considerable high value of overshoot. An important issue that should be noticed is that the system controlled with PID has constant magnitude for high frequency, which cause the system sensitive to high frequency noises and, consequently, instability. As a result, it can be said that the occupants' comfort is guaranteed when applying the proposed FPI controller. 

\section{Fractional-Order Reset Control}
\label{ADRSC}

Currently, reset control focuses on using structures which allow new resetting rules in order
to avoid limit cycle to be caused and improve the performance of the system.
This section presents two structures of reset control including fractional-order dynamics to avoid such a kind of problem: a fractional-order PI+CI controller and a general SISO reset controller, with fractional dynamics, with both fixed and variable resetting to non-zero values. It also gives some examples of application of these strategies. More details of these reset strategies can be found in \cite{HosseinNia2013t}.

\subsection{Fractional-order PI+CI controller}

\begin{figure}[ptbh]
\begin{center}
\includegraphics[width=0.375\textwidth]{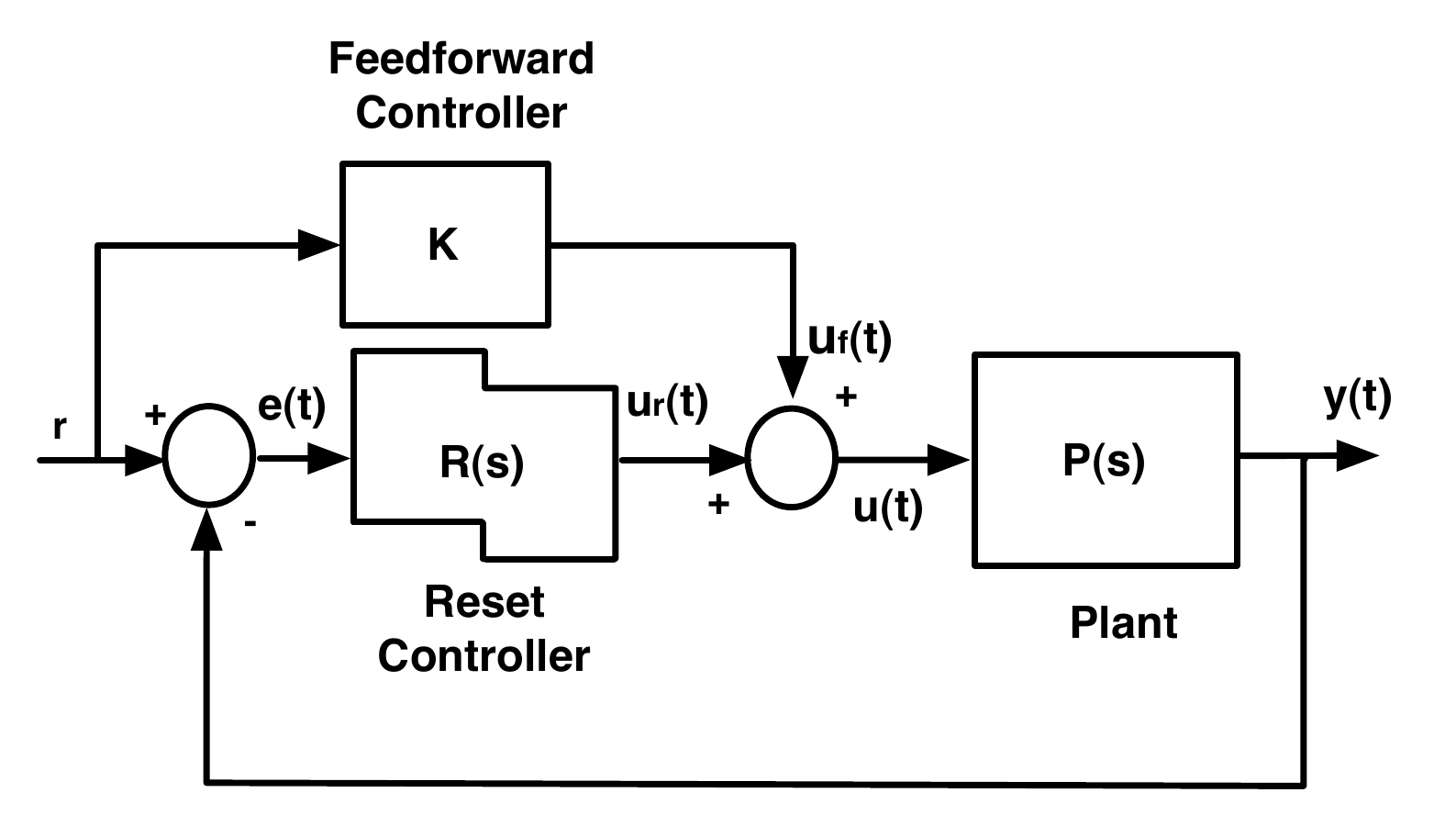}
\end{center}
\caption{Block diagram of a general reset control system}
\label{Reset_ff_Block_nC}
\end{figure}

It was demonstrated that, on the one hand, CI$^\alpha$ can increase the phase lag of the system \cite{HosseinNia2013b} and, on the other, PI+CI can be used to avoid limit cycles \cite{banos2011}. Therefore, a fractional-order PI+CI controller (both components of non-integer order, i.e., PI$^\alpha$ and CI$^\alpha$) can be given by:
\begin{equation}
R(s)=k_p\left(1+\frac{1-P_{reset}}{\tau_i s^\alpha}+\frac{P_{reset} \text{CI}^{\alpha}}{\tau_i} \right).
\label{controlFPICI}
\end{equation}
It can be written in state space with $A_r=0$, $B_r=\begin{bmatrix}1 & 1\end{bmatrix}^T$, $A_{R_r}=\begin{bmatrix} 1&0\\
0&0\end{bmatrix}$, $C_r=\begin{bmatrix} 0& \frac{k_p}{\tau_i} \end{bmatrix}$, $D_r=k_p$ (see equation (12) in part I of this work \cite{HosseinNia2014a} for more details). The describing function of PI$^\alpha$+CI$^\alpha$ is expressed as
\small
\begin{equation}
N(j\omega)=k_p\left(1+\frac{1-P_{reset}}{\tau_i(j\omega)^\alpha}+\frac{P_{reset}}{\frac{4\tau_i}{\pi\omega^\alpha}\left(\sin \left(\alpha\frac{ \pi}{2}\right)+\frac{\pi}{4}e^{-j\alpha \frac{\pi}{2}} \right)}\right).
\label{DF_FPICI}
\end{equation}
\normalsize

In \cite{HosseinNia2013t}, it has been shown that this controller allows to achieve both higher phase margin and crossover gain frequency than the base controller. Moreover, changing the order $\alpha$ in PI$^\alpha$+CI$^\alpha$, it is possible to obtain higher phase margin and crossover gain frequency than the PI+CI compensator when $P_{reset}=0.5$. At the same time, the lower the value of $\alpha$ and $P_{reset}$, the higher both the obtained phase margin and the crossover gain frequency. This means that a better performance in terms of both the speed of response and the relative stability can be obtained by means of PI$^\alpha$+CI$^\alpha$ compensator, overcoming limit cycle problem and improving the performance obtained with PI+CI.

\subsection{General reset control with fixed and variable reset}

A general fractional-order reset controller whose state is reset to $Kr$ when error crosses zero can be represented as
\begin{eqnarray}
\begin{matrix}
D^\alpha{x}_r(t)=A_r x_r(t)+B_re(t), \ e(t)\neq0,\\ 
x_r(t^+)=A_{R_r}x_r(t)+\frac{K}{n_{\mathcal{R}}c_r}B_{R_r}r, \ e(t)=0,\\
u_r(t)=C_rx_r(t)+D_re(t),
\end{matrix}
\label{resetnonzero}
\end{eqnarray}
where matrix $A_{R_r}  \in \mathbb{R}^{n_r\times n_r}$ identifies that subset of states $x_r(t)$ that are reset (the last ${\mathcal{R}}$ states) and has the form $A_{R_r}=\begin{bmatrix}
I_{n_{\bar{\mathcal{R}}}}& 0 \\ 
0 & 0_{n_{\mathcal{R}}}
\end{bmatrix}$, with $n_{\bar{\mathcal{R}}}=n_r-n_{\mathcal{R}}$ and $n_{\mathcal{R}}$ the length of the last ${\mathcal{R}}$ states, $B_{R_r}=\begin{bmatrix} 0 &1 \end{bmatrix}^T$, $C_r=c_r\begin{bmatrix} 0 & 1 \end{bmatrix}$, $c_r\in\mathbb{R}$. And $I$ and $0$ denote identity and zero matrices with proper dimension, respectively. 

Controller (\ref{resetnonzero}) is a reset control with feedforward where its feedforward part becomes active when error crosses zero at the first time. Actually, it activates the feedforward gain when it is necessary, which is the first reset time, in order to avoid limit cycles. Therefore, the general reset controller, unlike the reset controller with feedforward, maintains the same rise time as the base controller. This strategy was designed based on the reset controller with feedforward proposed in \cite{nesic2011}, in which the reset controller reset to a non-zero value that is inverse of DC gain of the system. 

Let us denote the transfer function of the base controller as $R_{base}(s)$. According to Fig. \ref{Reset_ff_Block_nC}, in presence of the error, the closed-loop transfer function of the system controlled by the reset controller with feedforward and general reset controller are, respectively, 
$ \frac{(K+R_{base}(s))P(s)}{1+R_{base}(s)P(s)}$ and $ \frac{R_{base}(s)P(s)}{1+R_{base}(s)P(s)}$.
Comparing these transfer functions with the transfer function of a classic controller (controller with no reset), it is obvious that only the general reset controller preserves some specification of the classic controller like rise time. 


Likewise, controller (\ref{resetnonzero}) can be reshaped to reset periodically when $t=t_k$, similarly to the reset control with  fixed reset instants $t_k$, which will lead us to another general reset controller as follows:
\begin{eqnarray}
\begin{matrix}
D^\alpha{x}_r(t)=A_r x_r(t)+B_re(t), \ t\neq t_k,\\ 
x_r(t^+)=A_{R_r}x_r(t)+B_{R_r}\left(\frac{Kr-D_re(t_k)}{n_{\mathcal{R}}c_r}\right), \ t=t_k,\\
u_r(t)=C_rx_r(t)+D_re(t).
\end{matrix}
\label{fracresetnonzero}
\end{eqnarray}
Due to the fact that reset happens periodically, and not necessarily when error is zero, it should take place to a variable non-zero value, which is function of both DC gain of the system and error.

\begin{example}
General reset control for a second-order system
\end{example}
Let us now consider the dynamics of a micro-actuator plant described by (\cite{zheng2007}):
\begin{equation}
\begin{matrix} 
\dot{x}_{p_1}(t)=x_{p_2}(t), \\
\dot{x}_{p_2} (t)=-a_1x_{p_1}(t)-a_2x_{p_2}(t)+bu(t)\\
y(t)=x_{p_1}(t) \end{matrix},
\label{sysimp}
\end{equation}
where $x_{p_1},\ x_{p_2}$ are position and velocity of the moving stage with $a_1 = 10^6$, $a_2 = 1810$, and $b = 3\times10^6$. This system can be also given by its transfer function $P(s) =\frac{b}{s^2+a_2s+a_1}$. This example firstly compares different strategies with zero crossing and, then, controllers with periodic reset for this system.

Consider a reset controller with a PI as base linear controller and a periodic reset action, so:
\begin{equation}
\begin{matrix} 
\dot{x}_{r}(t)=e(t),\ t \neq t_k \\
x_{r}(t^+) =E_1x_{p_1}(t)+E_2x_{p_2}(t)+Gr(t),\ t = t_k\\
u(t)=\frac{k_p}{\tau_i}x_{r}(t)+k_pe(t) \end{matrix} ,
\label{rcimp}
\end{equation}
with $k_p = 0.08$ and $\tau_i = \frac{8}{3}\times10^{-4}$. The optimal solution is given by the constant matrices $E_1 = -2.8 \times 10^{-4},\ E_2 =-6.8 \times 10^{-7}$, and $G = 0.0014$ \cite{zheng2007}. For the general controller, similar values were used with $\alpha=1$.

Indeed, general reset controller, reset controller with feedforward and reset control with  fixed reset instants $t_k$  (\ref{rcimp}) reset to non-zero values. In particular, reset control with fixed reset instants $t_k$ will reset to $\frac{k_p}{\tau_i}\left(E_1x_{p_1}+E_2x_{p_2}+Gr\right)+k_pe$. As time tends to infinity, the states $x_{p_1}$ and $x_{p_2}$ and error tend to $r$, $0$ and $0$, respectively. Therefore, the control signal $u_r$ tends to $\frac{k_p}{\tau_i}\left(E_1+G\right)r=0.336$, which is very close to the feedforward gain for the unit step input, i.e., $K=\frac{1}{P(0)}=0.333$. Likewise, reset control with fixed reset instants, resets when $t = t_k$ at each $1$ ms, whereas general reset controller and reset controller with feedforward reset when $e=0$. 

\begin{figure}[ptbh]
\begin{center}
\includegraphics[width=0.4\textwidth]{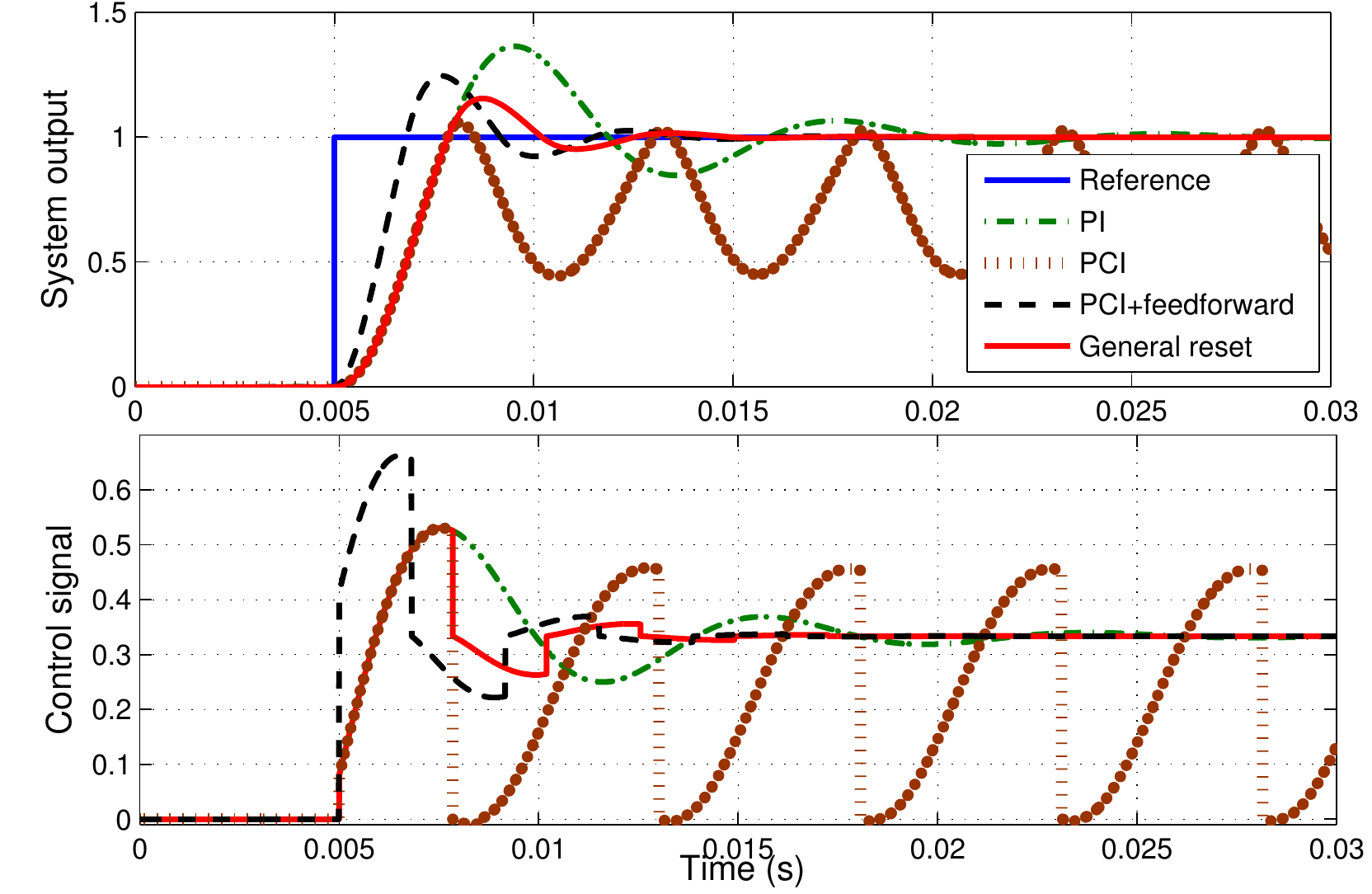}
\end{center}
\caption{Comparison of controllers with zero crossing reset for second-order system (\ref{sysimp})}
\label{improreset}
\end{figure}

\begin{figure}[ptbh]
\begin{center}
\includegraphics[width=0.4\textwidth]{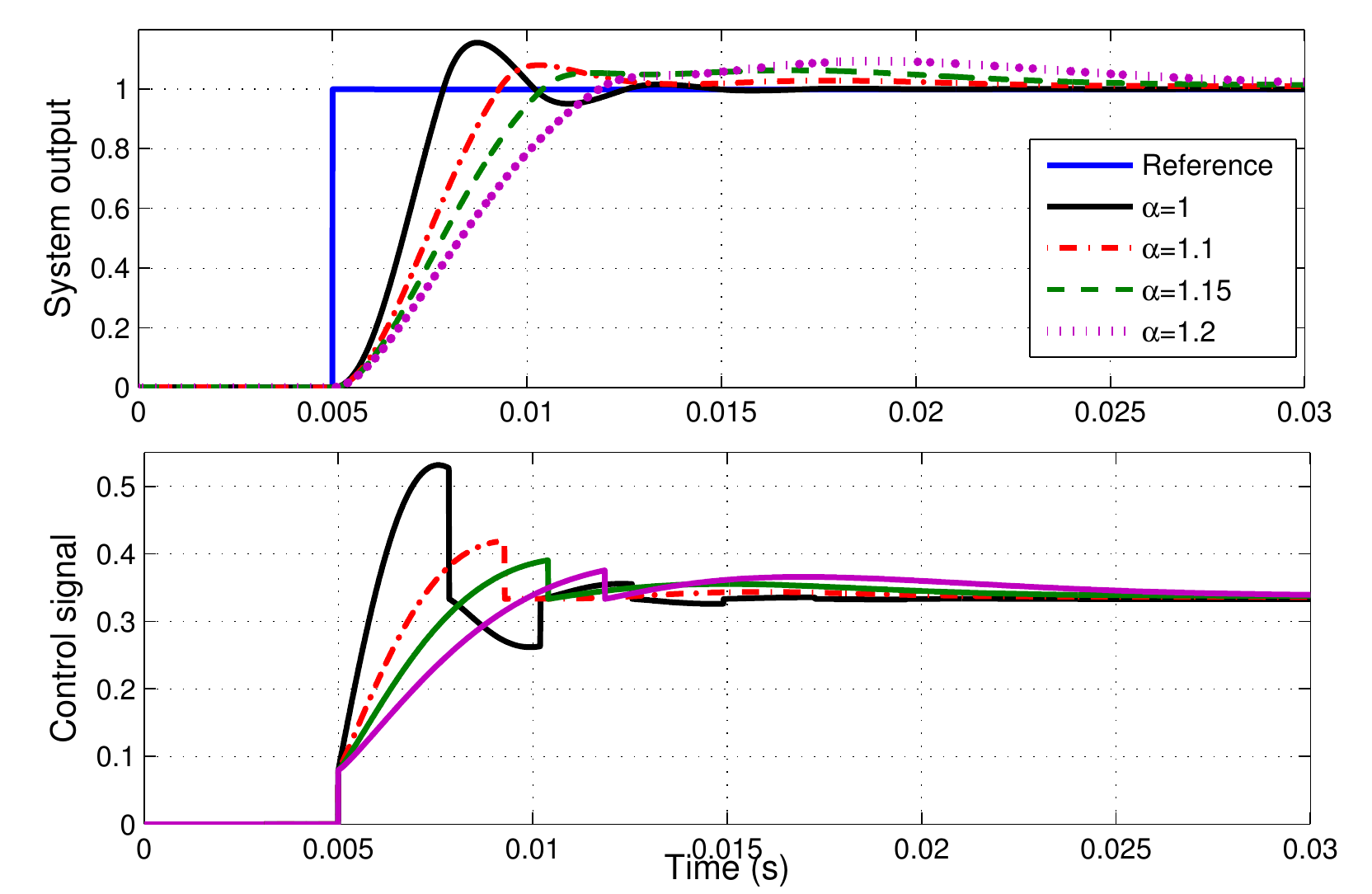}
\end{center}
\caption{Comparison of general reset control for different values of $\alpha$ for second-order system (\ref{sysimp})}
\label{frac_comp_reset}
\end{figure}

The step responses and control signals when applying reset controller with feedforward and general reset controller are shown in Fig. \ref{improreset}. The performance of the system using a PI and a PCI (PCI is a PI controller where the integrator is replaced by CI) were also obtained. As expected, reset controller with feedforward and general reset controller are able to eliminate the limit cycle caused by PCI, and this is because of the control signals reach the steady state value $K$.  Figure \ref{frac_comp_reset} compares general reset control for different values of $\alpha$. It can be seen that the higher the value of $\alpha$, the lower the overshoot but the slower the response. Thus, a trade off between an integer and a fractional-order general reset controller (in this case $\alpha=1.1$) may be a good way to overcome both limit cycle and overshoot at the same time. The feedforward gain in the reset controller with feedforward and the fractional-order CI in PCI$^{\alpha}$ cause different rise time in comparison with the classic PI controller. 

\begin{figure}[ptbh]
\begin{center}
\includegraphics[width=0.4\textwidth]{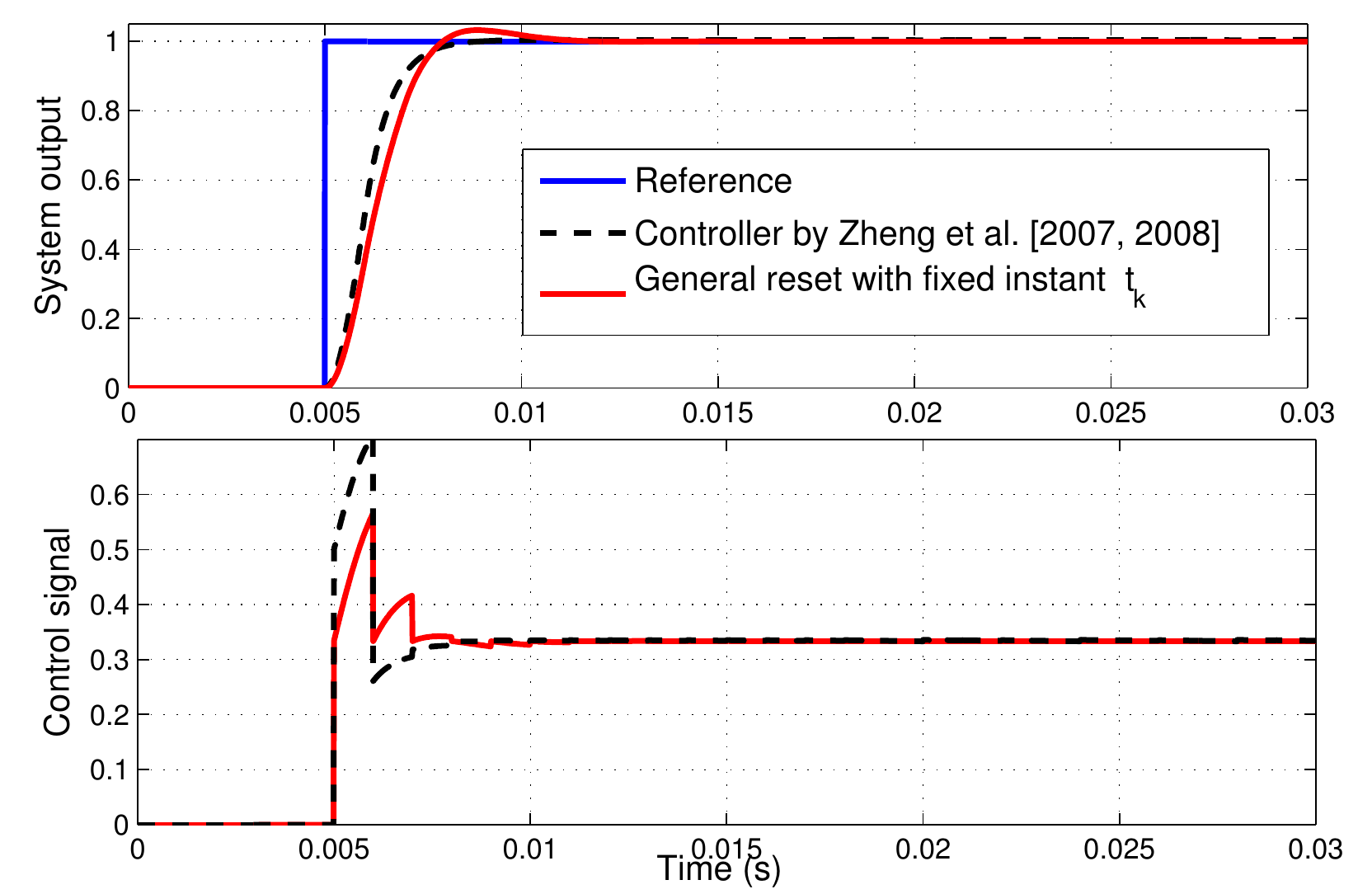}
\end{center}
\caption{Comparison of controllers with periodic reset for second-order system (\ref{sysimp})}
\label{compimproved}
\end{figure}

Now, consider a general reset control with periodic resetting with the following parameters: $\alpha=1$, $n_{\mathcal{R}}=1$, $A_r=0$, $B_r=1$, $c_r=C_r=\frac{k_p}{\tau_i}$ and $D_r=k_p$. Simulation results using this controller and reset controller with reset instants are depicted in Fig. \ref{compimproved} for $t_k=1$ ms. In comparison with the other strategies, it is seen that the overshoot is considerably reduced when applying controllers with periodic reset since they reset periodically before error reaches zero. However, it is worth mentioning that the general reset is capable of obtaining similar results than the controller proposed by \cite{zheng2007} but without an optimization process, making the design of the reset controllers simpler and more efficient. Notice that all the controllers have the same PI controller as base controller and, consequently, the system responses have similar rising time to the obtained with the classical PI controller, except with the PCI with feedforward controller. 

For comparison purposes, Table \ref{TABreset} gives the integral of the squared error (ISE), the maximum value of the control signal, the overshoot and the rising time for system (\ref{sysimp}) when applying the designed controllers. As observed, the application of periodically reset, in comparison with traditional zero crossing reset, reduces considerably the ISE and the overshoot, but changes the rising time --it is increased. Considering controllers with fixed reset instants, the system response, in terms of ISE and overshoot, is slightly better when applying the reset control in \cite{zheng2007}. However, as commented previously, the general reset control proposed in this work is easier to tune. On the other hand, among strategies with the classical reset condition, it is observed that the lowest value of the ISE is obtained when using the general reset controller. The worst result in terms of high control signal is obtained by the reset control with feedforward.
\small
\begin{table*}
\begin{center}
\caption{Performance of the designed controllers for second-order system (\ref{sysimp})}\label{TABreset}
{\small
\begin{tabular}{c||c|c|c|c|c|c}
& \multicolumn{4}{c|}{Strategies with zero crossing reset} & 
\multicolumn{2}{|c}{Strategies with fixed reset instants} \\ \hline
& PCI & PI & PCI+Feedforward & General reset & Controller by \cite{zheng2007} & General reset \\ 
\hline
ISE & 650.3390 & 31.5634 & 4.8773 & 3.1660 & 0.0870 & 0.1595 \\ \hline
Max(u) & 0.5315 & 0.5305 & 0.6639 & 0.5325 & 0.7054 & 0.5646 \\ \hline
$M_p$ ($\%$)  & 55.81 & 36.30 & 24.56 & 15.50 & 0.42 & 3.2 \\ \hline
$t_s$ (ms) & 0.284 & 0.284 & 0.18 & 0.284 & 0.403 & 0.30
\end{tabular}
}
\end{center}
\end{table*}
\normalsize

\begin{example} {Reset control for a velocity servomotor}
\end{example}

This example is to compare fractional- and integer-order reset strategies for the velocity control of a servomotor by \emph{Feedback} (see its description in \cite{feedback}), whose dynamic model is given by:
$P(s)=\frac{0.93}{0.61s+1}.$
Refer to \cite{HosseinNia13b} for more details of this application example.

Three base controllers, of integer- and fractional-order, i.e., a PI, a PID and a fractional-order PI (FPI), were tuned considering the following specifications related to phase margin $\phi_m$ and gain crossover frequency $\omega_{cg}$: $\phi_m\simeq45^\circ$ at $\omega_{cg}\simeq5.5$ rad/s (see their parameters in Table~\ref{TABRES}). Unfortunately, these base controllers can make the controlled system fast but very underdamped, so reset controllers are required to reduce overshoot and increase phase margin  (e.g. refer to \cite{banos2011,zheng2007}). Thus, replacing traditional integrators in the base controllers by CI or FCI, the following reset controllers were also obtained: a proportional CI (PCI), a proportional Clegg integro-differentiator (PCID) and a FPCI. It should be remarked that up to three design specifications can be fulfilled with the FPCI --there exists one more degree of freedom due to its order $\alpha$. Since only two specifications have to be fulfilled, the performance of the system was analyzed for different values of $\alpha$ for the FPCI --i.e., $0.5\leq\alpha\leq1$ with steps of $0.05$--, causing the following features on the system response: the higher the value of $\alpha$, the faster response and, on the contrary, the lower the value of $\alpha$, the less chance of limit cycle occurrence. Taking into account both issues, an intermediate value of $\alpha$ was chosen for the FPCI for this application: $\alpha =0.75$.
\begin{table}[pth]
\caption{Parameters of the base controllers}%
\label{TABRES}
\begin{center}
{\small
\begin{tabular}
[c]{l|c|c|c|c}
& $K_{p}$ & $K_{i}$ & $K_d$ & $\alpha$ \\\hline
PI & 1.6 &18.5 & - & - \\\hline
PID & 1.528 & 23.16 & 0.152 & - \\\hline
FPI & 0.067 & 13.4 & - & 0.75 \\
\end{tabular}
}
\end{center}
\end{table}

\begin{center}%
\begin{figure}[ptbh]
\begin{center}
\includegraphics[width=0.4\textwidth]{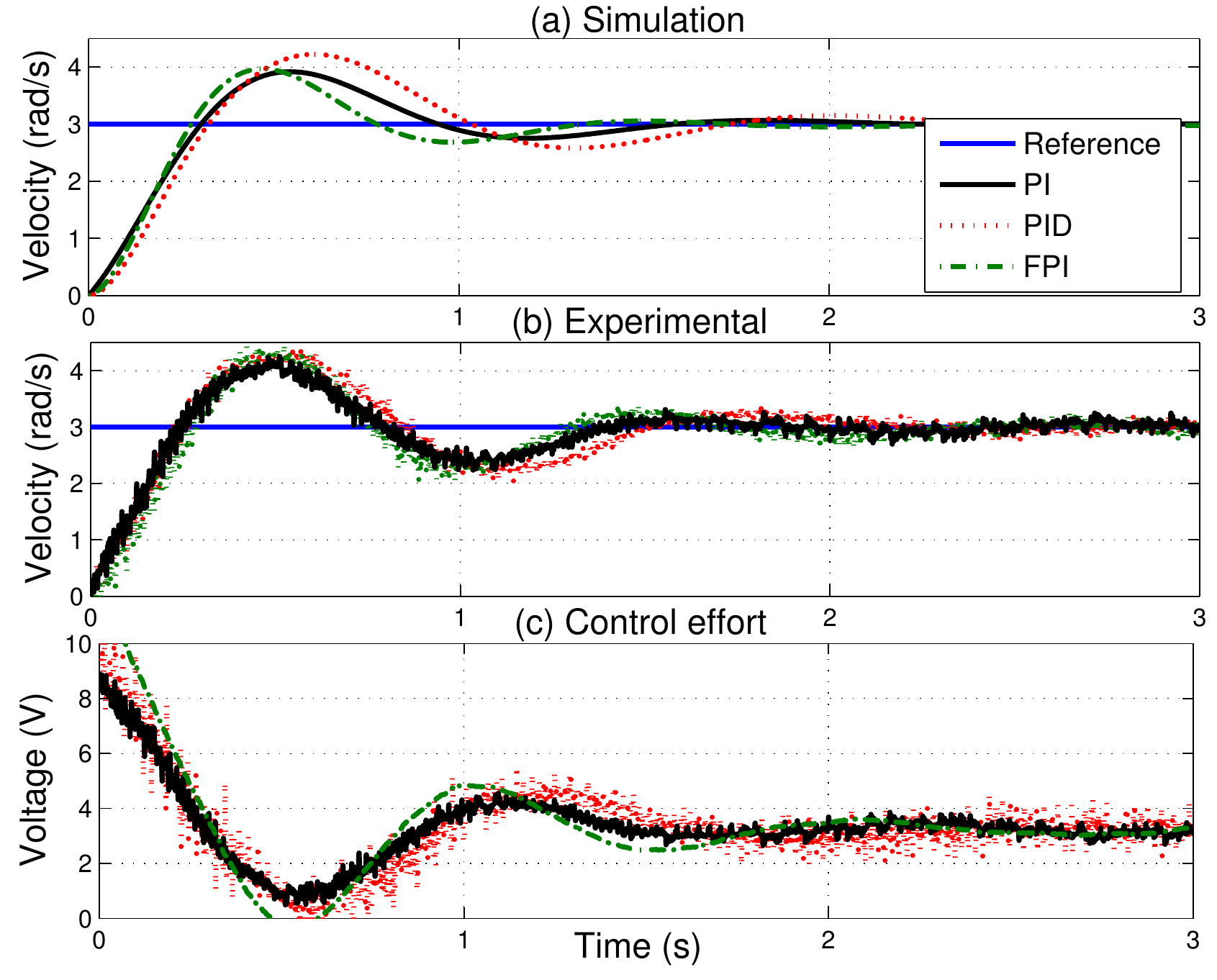}
\caption{Response of the servomotor  when applying the designed base controllers}%
\label{PracServo1}%
\end{center}
\end{figure}
\end{center}

\begin{center}%
\begin{figure}[ptbh]
\begin{center}
\includegraphics[width=0.4\textwidth]{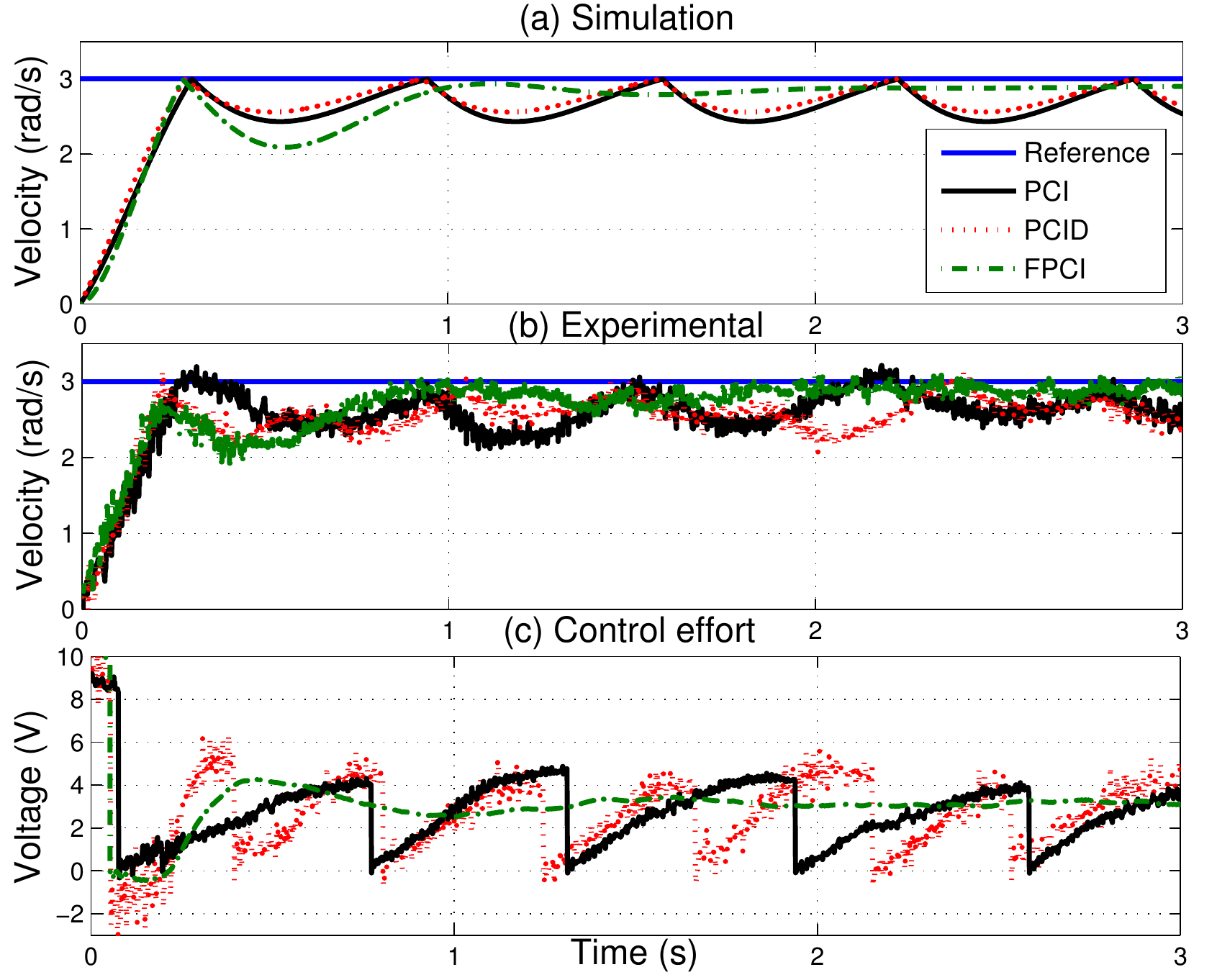}
\caption{Response of the servomotor when applying the designed reset controllers}%
\label{PracServo2}%
\end{center}
\end{figure}
\end{center}

The results obtained applying the base controllers are shown in Fig.~\ref{PracServo1}, where solid, dotted and dash-dotted lines refer to PI, PID and FPI, respectively. From this figure, it can be stated that: (i) the experimental results are similar to the simulated ones; (ii) all responses are stable but have a undesirable value of overshoot. Figure~\ref{PracServo2} shows the simulation and experimental results corresponding to the PCI, PCID and FPCI --solid, dotted and dash-dotted lines, respectively. As observed, simulation and experimental results are quite similar, as shown with the previous controllers. Moreover, the overshoot is reduced for all cases. It can be also seen that both the simulated and the experimental responses using PCI and PCID cause the occurrence of limit cycle. On the contrary, one can see that there is no such problem when applying the FPCI.
\small

\section{Conclusions}
\label{Conch2}
Part II of this paper was studied different fractional-order strategies involved in the control of hybrid systems (HS). In particular, two types of such techniques were reviewed, robust control of switching systems and reset control. Experimental and simulated examples were given to demonstrate their effectiveness. 

Since there is no a general agreement of the interpretation of state space representation of fractional-order systems, mainly concerning initial values (see e.g.~\cite{Sabatier13}), a further study should be carried out for fractional-order reset control taking into account this issue in future work.

\bibliographystyle{IEEEtran}

\begin{thebibliography}{10}
\providecommand{\url}[1]{#1}
\csname url@samestyle\endcsname
\providecommand{\newblock}{\relax}
\providecommand{\bibinfo}[2]{#2}
\providecommand{\BIBentrySTDinterwordspacing}{\spaceskip=0pt\relax}
\providecommand{\BIBentryALTinterwordstretchfactor}{4}
\providecommand{\BIBentryALTinterwordspacing}{\spaceskip=\fontdimen2\font plus
\BIBentryALTinterwordstretchfactor\fontdimen3\font minus
  \fontdimen4\font\relax}
\providecommand{\BIBforeignlanguage}[2]{{%
\expandafter\ifx\csname l@#1\endcsname\relax
\typeout{** WARNING: IEEEtran.bst: No hyphenation pattern has been}%
\typeout{** loaded for the language `#1'. Using the pattern for}%
\typeout{** the default language instead.}%
\else
\language=\csname l@#1\endcsname
\fi
#2}}
\providecommand{\BIBdecl}{\relax}
\BIBdecl

\bibitem{Gollu_89}
A.~Gollu and P.~Varaiya, ``Hybrid dynamical systems,'' in \emph{Proceedings of
  the 28th IEEE Conference on Decision and Control}.\hskip 1em plus 0.5em minus
  0.4em\relax IEEE, 1989, pp. 2708--2712.

\bibitem{schumacher_99}
A.~J. van~der Schaft and J.~M. Schumacher, \emph{Introduction to hybrid
  dynamical systems}.\hskip 1em plus 0.5em minus 0.4em\relax Springer-Verlag,
  1999.

\bibitem{Goebel_09}
R.~Goebel, R.~Sanfelice, and A.~Teel, ``Hybrid dynamical systems,''
  \emph{Control Systems Magazine, IEEE}, vol.~29, no.~2, pp. 28--93, 2009.

\bibitem{banos2011}
A.~Ba\~nos and A.~Barreiro, \emph{Reset Control Systems}.\hskip 1em plus 0.5em
  minus 0.4em\relax Springer Verlag, 2011.

\bibitem{Schutter2009}
B.~D. Schutter, W.~Heemels, J.~Lunze, and C.~Prieur, \emph{Handbook of Hybrid
  Systems Control--Theory, Tools, Applications}.\hskip 1em plus 0.5em minus
  0.4em\relax Cambridge University Press, 2009, pp. 31--35.

\bibitem{Monje10}
C.~A. Monje, Y.~Q. Chen, B.~M. Vinagre, D.~Xue, and V.~Feliu,
  \emph{Fractional-order Systems and Controls. Fundamentals and
  Applications}.\hskip 1em plus 0.5em minus 0.4em\relax Springer, 2010.

\bibitem{Podlubny_99a}
I.~Podlubny, \emph{Fractional Differential Equations. An Introduction to
  Fractional Derivatives, Fractional Differential Equations, Some Methods of
  Their Solution and Some of Their Applications}.\hskip 1em plus 0.5em minus
  0.4em\relax Academic Press, San Diego - New York - London, 1999.

\bibitem{HosseinNia2014a}
S.~H. HosseinNia, I.~Tejado, and B.~M. Vinagre, ``Hybrid systems and control
  with fractional dynamics ({I}): {M}odeling and analysis,'' in
  \emph{Proceedings of the 2014 International Conference in Fractional
  Differentiation and its Applications (ICFDA'14)}, 2014.

\bibitem{Hosseinnia2013}
------, ``A method for the design of robust controllers ensuring the quadratic
  stability for switching systems,'' \emph{Journal of Vibration and Control},
  2013.

\bibitem{Monje08}
C.~A. Monje, B.~M. Vinagre, V.~Feliu, and Y.~Q. Chen, ``Tuning and auto-tuning
  of fractional order controllers for industry applications,'' \emph{Control
  Engineering Practice}, vol.~16, no.~7, pp. 798--812, 2008.

\bibitem{HosseinNia2011}
S.~H. HosseinNia, I.~Tejado, B.~M. Vinagre, V.~Milan\'{e}s, and
  J.~Villagr\'{a}, ``Low speed control of an autonomous vehicle using a hybrid
  fractional order controller,'' in \emph{Proceedings of the 2nd International
  Conference on Control Instrumentation and Automation (ICCIA'11)}, 2011.

\bibitem{HosseinNia_12}
------, ``Experimental application of hybrid fractional order adaptive cruise
  control at low speed,'' \emph{IEEE Transactions on Control Systems
  Technology}, 2014, in Press.

\bibitem{HosseinNia2013t}
S.~H. HosseinNia, ``Fractional hybrid control systems: Modeling, analysis and
  applications to mobile robotics and mechatronics,'' Ph.D. dissertation,
  University of Extremadura, 2013.

\bibitem{HosseinNia2013b}
S.~H. HosseinNia, I.~Tejado, and B.~M. Vinagre, ``Stability of fractional order
  switching systems,'' \emph{Computer \& Mathematics with Applications},
  vol.~66, no.~5, pp. 585--596, 2013.

\bibitem{nesic2011}
D.~Nesic, A.~R. Teel, and L.~Zaccarian, ``Stability and performance of {SISO}
  control systems with first-order reset elements,'' \emph{IEEE Transactions on
  Automatic Control}, vol.~56, no.~11, pp. 2567--2582, 2011.

\bibitem{zheng2007}
J.~Zheng, Y.~Guo, M.~Fu, Y.~Wang, and L.~Xie, ``Improved reset control design
  for a {PZT} positioning stage,'' in \emph{Proceedings of the IEEE
  International Conference on Control Applications}, 2007, pp. 1272--1277.

\bibitem{feedback}
\emph{Analogue Servo. Fundamentals Trainer 33-001}, Feedback Instruments Ltd.,
  UK.

\bibitem{HosseinNia13b}
S.~H. HosseinNia, I.~Tejado, and B.~M. Vinagre, ``Fractional-order reset
  control: {A}pplication to a servomotor,'' \emph{Mechatronics}, vol.~23,
  no.~7, pp. 781--788, 2013.

\bibitem{Sabatier13}
J.~Sabatier, C.~Farges, and J.-C. Trigeassou, ``Fractional systems state space
  description: Some wrong ideas and proposed solutions,'' \emph{Journal of
  Vibration and Control}, 2013.

\end{thebibliography}

\end{document}